
\magnification=\magstep1
\input epsf
\hsize 6.5truein 
\vsize 8.9truein 
\baselineskip 19pt plus 2pt minus 1pt 

\vskip 1in
%
\hbadness=10000
\newbox\refsize
\setbox\refsize\hbox to\hsize{ }
\hbadness=1000
\def\endpage{\par\vfill\eject}
\newcount\pointnum
\pointnum = 0
\def\pointbegin{\pointnum = 0 \point }
\def\point{\par \global\advance\pointnum by 1
\noindent\hangindent .08\wd\refsize \hbox to .08\wd\refsize{\hfill
\the\pointnum .\quad}\unskip}
\newdimen\offdimen
\def\offset#1#2{\offdimen #1
   \noindent \hangindent \offdimen
   \hbox to \offdimen{#2\hfil}\ignorespaces}
\def\endlist{\par}

%

\hfuzz=5pt
\baselineskip 12pt plus 2pt minus 2pt
\vskip 24pt
\centerline {\bf Scaling and Selection in Cellular Structures and Living
Polymers}
\vskip 36pt
\centerline {David Mukamel}
\vskip 12pt
\centerline {\it Department of Physics of Complex Systems, The Weizmann
Institute of Science}
\centerline {\it Rehovot 76100, Israel}
\vskip 60pt
\centerline{\bf ABSTRACT}
\smallskip
The dynamical behavior of two types of non-equilibrium systems is
discussed: $(a)$ two-dimensional cellular structures, and $(b)$ living
polymers. Simple models governing their evolution are introduced and steady
state distributions (cell side in the case of cellular structures and
length in the case of living polymers) are calculated. In both cases the
models possess a one parameter family of steady state distributions.
Selection mechanism by which a particular distribution is dynamically
selected is discussed. \baselineskip 18pt plus 4pt minus 4pt \vskip 0.5in
\noindent PACS No: 05.70.Ln 47.20.Ky 82.70.Rr \medskip \vfill
\break
\baselineskip=14pt plus 3pt minus 2pt
\vskip0pt plus.1\vsize\penalty-250
\vskip0pt plus-.1\vsize\vskip24pt plus12pt minus6pt

Pattern selection is well known to take place in systems far from thermal
equilibrium such as growing crystals, reaction diffusion problems and in
many others. In these systems a particular pattern, or structure, is
dynamically selected out of a family of possible structures. The canonical
problem for which selection has been studied and explicitly demonstrated is
that of a front propagation, where the velocity with which a stable state
propagates into an unstable one is selected $[1,2]$. This mechanism was
then applied to study the pattern selection occuring in systems quenched
beyond their limit of stability $[3-8]$.

In the present paper we briefly review two interesting classes of systems
which have recently been suggested to display dynamics governed by
selection. The first class includes systems with cellular structures
$[9,10]$ and the second is living polymers $[11]$. Here the selected
quantity is not a spatial structure but rather the steady state
distributions of certain quantities like that of the cell side in the case
of cellular structures and the polymer length distribution in the case of
living polymers. This review summarizes results obtained in Refs. $[9-11]$.
\bigskip
\noindent
{\bf A. Two Dimensional Cellular Structures} \medskip
Cellular structures are rather common in nature $[12]$. Examples include
polycrystal $[13-15]$, foams, soap froths $[16,17]$, magnetic bubbles
$[18-21]$, and many others. In many
cases these structures are metastable and they keep changing in time,
evolving towards equilibrium. For example in the case of polycrystals, the
cells are composed of single crystallites which are randomly oriented.
Neighboring cells are separated by grain boundaries which cost some energy.
In an attempt to reduce this energy the cells coarsen, thus reducing their
area and the energy of the grain boundaries. Similarly, foams are made up
of liquid film membranes separating gas filled cells. The energy of the
system is given by the surface energy of the membranes. Here too the energy
is reduced via the coarsening process. The main difference between the two
systems is that while the surface free energy of liquid membranes is
isotropic, the grain boundary energy strongly depends on the relative
orientation of the neighboring crystallites and on the direction of the
boundary. This certainly affects the details of the dynamics, but not the
general tendency of coarsening. The evolution of these systems leads to a
scale invariant state, with steady state distributions of some of their
properties. Most studies so far have concentrated on two dimensional $(2D)$
systems which are much simpler to investigate both theoretically and
experimentally than in $3D$ $[22-29]$. Here we discuss the evolution of
$2D$ soap froth.

Consider a two dimensional array of soap bubbles made up of liquid film
membranes separating gas filled cells. The edges of the cells meet at
vertices with coordination number $q=3$. The reason is that any vertex with
higher coordination number tends to split into several vertices,
generically reducing the energy, or the total edge length, of the
configuration. If the gas cannot diffuse through the membranes, the system
reaches a mechanical equilibrium in which the angle between any pair of
edges meeting at a vertex is $120^\circ $. However, due to the permeability
of the membranes, the gas diffuse from one cell to another and the system
slowly evolves in a way such that its total edge length is reduced. In 1952
von Neumann considered the evolution of $2D$ soap froth under the
assumption that the diffusion process is sufficiently slow so that the
system is at a mechanical equilibrium at any given time. He showed that in
this limit, the area of an $l$ sided cell, $A_l$, satisfies the following
equation $[22]$
$$
{d A_l \over dt} = K (l-6)
\eqno (1)
$$
where $K$ is a constant which is proportional to the surface tension and to
the permeability of the membranes.
This is a very important and surprising result, since it shows that the
evolution of the area of a cell is independent of the details of its
neighborhood (such as the area, the shape and the number of edges of the
adjacent cells), but only on the number of edges of the cell itself. Eq.
(1) suggests that $l < 6$ sided cells shrink and eventually disappear while
cells with $l > 6$ sides grow in time. The area of hexagonal cells remains
unchanged. As a result of the disappearance of an $l < 6$ cell, the number
of sides of its neighboring cells change (see Fig.1a). When the
disappearing cell is a triangle, each of its adjacent cells loose a side.
For a disappearing square, two of its neighbors loose a side and two remain
with the same number of sides. In the case of a pentagon, two neighbors
loose a side, one gain a side and two remain unchanged. These events are
known as $T2$ processes. As a result of these processes, $l< 6$ sided cells
do not disappear from the system altogether. This is consistent with
Euler's law which states that $V-E+F=2$ where $V$ is the number of
vertices, $E$ is the number of edges and $F$ is the number of faces of a
closed two-dimensional graph in a plane. Using the fact that in our case
three edges meet at each vertex, it is easy to show that the the average
number of sides per cell is 6. Therefore, any configuration in which $l>6$
sided cells are present, must include $l<6$ sided cell too. In addition to
the $T2$ processes, one also finds $T1$ processes corresponding to edge
switching events (see Fig.1b). In soap froth these are rather
rare events and therefore we will not consider them here. Also, two sided
cells are metastable and they are not generated during the evolution. We
therefore assume that the system is composed of $l \geq 3$ sided cells
only.

\medskip
\centerline{\epsfysize5truein\epsfbox{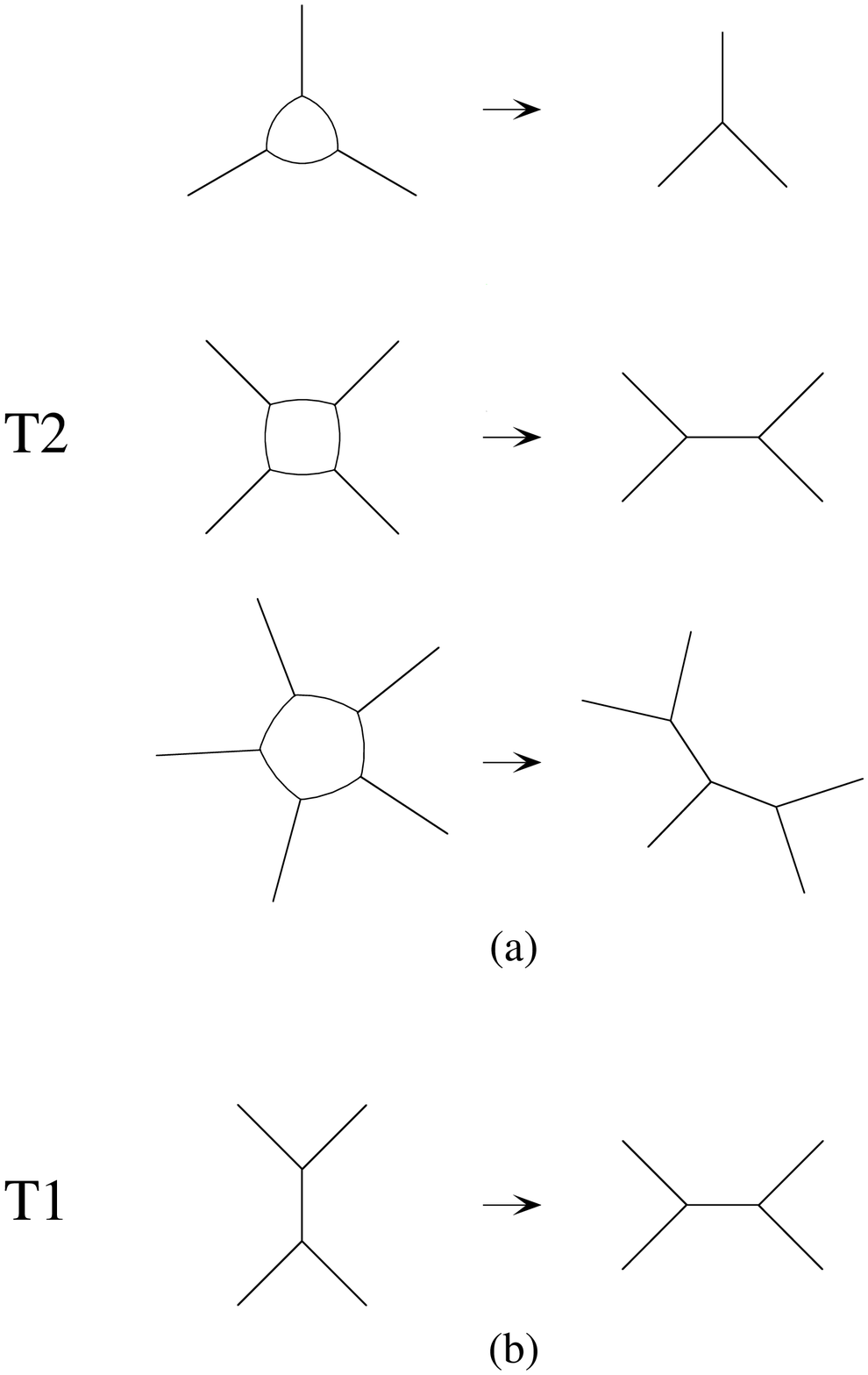}}
\nobreak\smallskip\nobreak
\offset{70pt}{\bf Figure 1:}
$(a)$ T2 processes, describing the disappearance of 3, 4 and 5 sided cells,
respectively. $(b)$ side switching T1 process.
\bigskip

To study the evolution of soap froth we consider $N_l(t)$, the number of
$l$- cells (i.e. $l$- sided cells) at time $t$. Let $N= \sum_{l=3}^\infty
N_l$ be the total number of cells in the system, and define $$ x_l= N_l /N
\eqno (2)
$$
Clearly, the concentration $x_l$ of $l$- cells satisfy $$
\sum _{l=3}^\infty x_l=1
\eqno (3a)
$$
Moreover, due to Euler's law one also has $$
\sum _{l=3}^\infty l x_l =6
\eqno (3b)
$$
To proceed, we introduce three parameters $w_3, w_4$ and $w_5$ which yield
the vanishing rates of triangles, squares and pentagons, respectively.
These rates together with the $T2$ processes determine the evolution
equations of $N_l$. For example the resulting equation for $N_3$ takes the
form $$ {d N_3 \over d t} = -w_3
N_3+ [3 P_R(4,3) w_3 N_3 + 2 P_R(4,4) w_4 N_4 +2 P_R(4,5) w_5 N_5] -
P_A(3,5) w_5 N_5
\eqno (4)
$$
where $P_R(l,m)$ is the probability that an edge of a disappearing $m \leq
5$-cell is shared by an adjacent $l$-cell which undergoes a side removal
process. Similarly, $P_A(l,5)$ is the probability that an edge of a
disappearing pentagon is shared by an adjacent $l$- cell which undergoes a
side addition process. In this equation the first term represents the rate
at which triangles disappear due to the von-Neumann evolution. The next
three terms yield the rate at which squares turn into triangles due to the
disappearance of a neighboring $3,4$ and $5$ cells, respectively. The last
term gives the rate at which triangles gain a side and become squares due
to the disappearance of a neighboring pentagon.

We now introduce the mean field approximation for $P_R(l,m)$ and
$P_A(l,5)$. In considering $P_A(l,5)$ we neglect all correlations between
neighboring cells and take
$$
P_A(l,5)={ l N_l \over {6 N}} = {{l x_l} \over 6} \eqno(5) $$
In $P_R(l,m)$ some correlations have to be retained. In particular, in
order not to
generate $2$-cells a side removal process should not take place in a
triangle. We
therefore take
$$
P_R(l,m)=\left\{\matrix{0 \quad &, & \quad l=3 \cr l x_l /(6-3x_3) \quad &,
& \quad
l>3 \cr}\right. \eqno(6)
$$
where the denominator $(6-3 x_3)$ is taken to ensure that the normalization
condition $\sum _{l=3}^\infty P_R(l,m) =1$ is satisfied. One can easily
write down similar equations for
$N_l$,
$l >3$. Summing these equations one obtains an equation for the decreasing
rate of the
total number of cells $$ {d N \over d t} = -(w_3 N_3 + w_4 N_4 +w_5 N_5)
\eqno(7)
$$
Combining this equation with the equations for $N_l$ one obtains the
following equations for $x_l$, $l \ge 3$ $$
{d x_l \over d t} =-w_l x_l + K_0 x_l
+K_1[(l+1) x_{l+1} -\alpha _l l x_l]
- K_2 [l x_l - \alpha _l (l-1) x_{l-1}]
\eqno(8)
$$
where
$$\eqalign{K_0 & = w_3 x_3 +w_4 x_4 + w_5 x_5\cr K_1 & = (3 w_3 x_3 + 2 w_4
x_4 +2 w_5 x_5) / (6-3 x_3)\cr K_2 &= w_5 x_5 / 6\cr}\eqno(9)
$$
and
$$
\alpha _l = \left\{\matrix{0\quad ,&\quad l=3\cr 1 \quad , & \quad
l>3\cr}\right.
$$
Also $w_l=0$ for $l >5$.

We are interested in the steady state solutions of Eqs. (8,9) for which $d
x_l / d t =0$. It is easy to demonstrate that these equations have a one
parameter
family of steady state solutions. To see that this indeed is the case we
notice that
for any given $0 < x_3 <1 $ the first two equations (corresponding to
$l=3,4$) could
be solved to yield $x_4$ and $x_5$. One can then solve iteratively for the
rest of
the distribution, finding $x_l$ from the $(l-1)$-th equation. The family of
distributions obtained in this way may thus be parametrized by $x_3$.

We now consider the steady state distributions in more detail. Let $$
y_l =l x_l
\eqno(10)
$$
The steady state equations for $l>5$ take the form $$ K_1 y_{l+1} - (K_1 +
K_2 -{K_0 \over l}) y_l + K_2 y_{l-1}=0 \eqno(11) $$
These equations have solutions of the form $$ y_l = \lambda ^l l^{\kappa}
(1 + O({1 \over l})) \eqno(12) $$
Inserting this form into Eq. (11) the parameters $\lambda$ and $\kappa$ may be
obtained. Two solutions are found. One with
$$
\lambda_1={K_2 \over K_1} ,~~ { \kappa_1={K_0 \over {K_1-K_2}}} \eqno (13a) $$
and the other with
$$
\lambda_2=1 ,~~ {\kappa_2={-K_0 \over {K_1-K_2}}} \eqno(13b) $$
The general solution, therefore, takes the form $$
x_l=A\left({K_2\over K_1}\right)^ll^{\kappa _1-1}+Bl^{\kappa_2-1} \eqno(14)
$$
The full distribution (including the $l \le 5$ concentrations) is found by
first
chosing a particular value for $x_3$ and solving the $l=3,4$ equations to
obtain $x_4$ and $x_5$. One then uses the form (14) for the $l>5$ cell
concentrations and
determine the parameters $A$ and $B$ by demanding that the two sum rules
($3a$ and
$3b$) are satisfied. The tail of the distribution is thus given as a sum of
two terms: an $A$ term which decays exponentially with $l$ and a $B$ term
which decays
algebraically. By varying $x_3$ the whole family of possible steady state
distributions is obtained.

In calculating these distributions it is found that $B$ is positive
for sufficiently small $x_3$, and it changes sign at some point $x_3 =
x_3^*$. The distributions with $B<0$ are unphysical since $x_l$ become
negative for large $l$. Moreover, linear stability analysis shows that
those solutions with $B<0$ are unstable while those with $B > 0$ are
stable. As in other problems where selection takes place it is expected
that the physical solution which exhibits the fastest decay of $x_l$
with $l$ is the one which is selected. Here it is the $B=0$ solution,
which is the only one exhibiting exponential decay. All other
physically relevant distributions (with $B>0$) display algebraic decay for
large $l$.
Numerical studies of the  dynamical equations indicate that indeed this is
the case
and that the system evolves towards the $B=0$ steady state distribution.
The resulting distribution agrees well with experimental results obtained
in soap froth $[9,10]$.

In order to check how good is the mean field approximation applied in this
work,
we consider now the case in which only $T1$ processes take place. Here it
is assumed
that the gas does not diffuse from one cell to another, hence $T2$
processes do not
take place and cells do not disappear. Thus, the number of cells in the
system is
conserved throughout the evolution. This problem has been solved exactly
$[30]$,
yielding a
distribution which for large $l$ takes the form $$ x_l\sim \lambda ^l/\sqrt{l}
\eqno(15)
$$
with $\lambda= 3/4$.

To model this process within the mean field approximation, we define the
rate $w$ as the probability that a $T1$ process takes place per time unit
per edge. Each such event results in a removal of a side from two cells and
addition of a side to two others. The process cannot take place if either
of the two neighboring cells which lose an edge in this process is a
three-cell. This ensures that no two-cells are generated in this dynamics.
The evolution equations of $x_l$ are given by $$
{d x_l \over dt} =w{6-3x_3\over 6}[(l+1)x_{l+1}-lx_l]+w\left({6-3x_3\over
6}\right)^2[(l-1)x_{l-1}-lx_l]
\eqno(16)
$$
These equations process a single fixed point distribution $$
x_l\sim \lambda^l/l
\eqno(17)
$$
with $\lambda \approx 0.85$. This qualitatively agree with te exact answer.
\bigskip
\noindent
{\bf B. Living Polymers}
\medskip
Another example of a dynamical system whose steady state distribution is
uniquely selected out of a family of distributions is given by living
polymers. Here we consider a system of aggregates (say, polymers or
micells), each composed of $n$ units. Let $c_n$ be the number of $n$- mers
in the system. Under equilibrium condition the system reaches a well
defined length distribution $c_n$. We consider the case in which the system
is driven out of equilibrium by pulling out monomers (or small size
aggregates) at some rate. Clearly, $c_n$ decrease with time and the system
eventually vanishes. The question is how does the length distribution $c_n$
look like in the long time limit. To study this problem we consider the
concentrations $x_n = c_n/c$ where
$$
c= \sum_{n=1}^\infty n c_n
\eqno (18)
$$
and ask whether these quantities reach a well defined limit for large time.

As in the case of cellular structures, one finds that the dynamical
equations which govern the evolution of $x_n$ possess not just a single
fixed point but rather a one parameter family of steady state
distributions. Again a dynamical selection takes place by which a
particular distribution is selected. To study this behavior we introduce a
model describing the evolution of living polymers. For simplicity we assume
that the interaction between the various $n$-mers take place via
association-dissociation process in which an $n$-mer absorbs or emits a
single monomer and becomes $(n+1)$ or $(n-1)$-mer, respectively. The
equations governing this process take the form: $$ \eqalign
{
{d c_1 \over dt} &= - k x_1 \sum_{n=1}^{\infty} c_n + \bar k
\sum_{n=2}^{\infty}c_n -a c_1, \cr {d c_2 \over dt} &= k x_1( c_1/2 -c_2) +
\bar k (c_3-c_2/2), \cr {d c_n \over dt} &= k x_1 (c_{n-1} -c_n ) +\bar k
(c_{n+1} -c_n) \quad {\rm for}\quad n \ge 3. \cr } \eqno(19)
$$
where $\bar k$ is the dissociation rate, $k x_1$ is the association rate of
an $n$-mer with a monomer and $a$ is the rate at which monomers are pulled
out of the system, $x_1$. The association rate is taken to be proportional
to the concentration of the monomers in the system. For simplicity it is
assumed that $k$ and $\bar k$ are independent of $n$. This assumption does
not change the qualitative results derived here. The equation for $c_2$ is
somewhat different from equations for $c_n$ with $n>2$. First, the rate at
which two monomers combine to yield a dimer is $k x_1 /2$ and not $k x_1$.
In addition, while a dissociation of $n$-mer with $n>2$ can take place near
either of its two ends, in the case of a dimer the dissociation can take
place only at one point. Hence the factor of a $1/2$ in its dissociation
rate. Summing Eqs.(19) one finds
$$
{d c \over dt} = -ac_1 ,
\eqno(20)
$$
This equation yields the rate at
which the number of units in the system decreases with time.

Combining Eqs.(20) with (19) one obtains the following equations for $x_n$. $$
\eqalign
{
{d x_1 \over dt} &=-k x_1 \sum_{n=1}^{\infty} x_n + \bar k
\sum_{n=2}^{\infty} x_n -a x_1 + a x_1^2 ,\cr {d x_2 \over dt} &= k x_1(
x_1/2 -x_2) + \bar k (x_3 -x_2/2) + a x_1 x_2, \cr {d x_n \over dt} &= k
x_1 (x_{n-1} -x_n) + \bar k (x_{n+1} -x_n) + a x_1 x_n \quad {\rm for}
\quad
n \ge 3. \cr
} \eqno (21 )
$$
This is a set of non-linear equations with non-local interactions of the
type found in the equations describing the evolution of cellular
structures. Here the "head" of the distribution, $x_1$, directly interacts
with its "tail", $x_n$, for arbitrarily large $n$. The fixed point
distribution of these equations yield the long time behavior of $c_n$. Note
that the normalization
$$
\sum_{n=1}^{\infty} n x_n =1 \ . \ \eqno (22) $$ is preserved by Eqs. (21).

Eqs. (21) possess a one parameter family of fixed point distributions. To
see that this is the case one first makes an arbitrary choice for $x_1$ and
$x_2$. The second equation of (21) determines $x_3$. One can then
successively solve the remaining equations, obtaining $x_n$ for $n>3$. The
two free parameters $x_1$ and $x_2$ may now be varied so as to satisfy the
equation for $x_1$ (or equivalently, the normalization condition (22)),
leaving one free parameter, say, $x_1$.

The fixed point distributions may easily be obtained by noting that for a
given $x_1$, Eqs. (21) are linear in $x_n$, $n \ge 2$. Setting $d x_n /dt
=0$ one finds that the fixed point distributions are of the form $$
x_n = A \lambda_-^{n-2} + B \lambda_+^{n-2}, \quad n\ge 2, \eqno (23) $$
where $\lambda_\pm$ are the roots of the characteristic equation $$
{\bar k \lambda^2 -(\bar k +k x_1 -a x_1) \lambda + k x_1 = 0}. \eqno (24) $$
Here $A$ and $B$ are parameters which are determined by the first two
equations in (22)
$$
\eqalign{
{A \over {1-\lambda_-}}+{B \over {1-\lambda_+}}&={{(k-a)x_1^2+ax_1} \over
{\bar k-kx_1}},\cr \cr
A \lbrack \bar k \lambda_- -(\bar k/2 +k x_1 - a x_1) \rbrack &+ B \lbrack
\bar k \lambda_+ -(\bar k/2 +k x_1 -a x_1) \rbrack =-k x_1^2 /2 \ . \ \cr}
\eqno (25) $$

Eqs. (25) yield the two parameters $A$ and $B$, thus determining the length
distribution. Clearly, all $x_n$'s obtained in this way must be
non-negative for the distribution to be physically meaningful. For this to
be the case one has to require that (a) the roots $\lambda_{\pm}$ are real
with $0 < \lambda_- < \lambda_+ <1$, (b) $x_2 =A+B >0$ and (c) $B>0$. These
requirements ensure that $x_n$ decays with $n$ purely exponentially without
oscillations. When the decay is oscillatory some of the $x_n$'s become
negative and thus unphysical. The $B>0$ requirement is needed since $B$
corresponds to the slower decay rate of $x_n$. A negative $B$ yields
negative $x_n$'s for large $n$.

We now consider the question of selection of a unique steady state
distribution out of this family. By examining Eq. (24) one can verify that
for any set of parameters $k, \bar k $ and $a$ there exists an $x_M$ such
that as long as $0 < x_1 < x_M$, condition $(a)$ is satisfied. It is also
easy to verify by solving Eq. (25) that condition $(b)$ is satisfied as
long as $0 < x_1 < x_M$. As for condition $(c)$, one can show that the
parameter $B$ is positive for small $x_1$ and changes sign at $x_1=x_S$,
where $x_S$ depends on the parameters $k, \bar k $ and $a$. For $x_S > x_M$
all fixed points with $0 < x_1 < x_M$ are physically relevant. On the other
hand for $x_S < x_M$ only those with $0 < x_1 < x_S$ correspond to physical
distributions. These two cases yield two different selection mechanisms.
Following the work of Aronson and Weinberger $[1]$, and Dee {\it et al}
$[3]$ one may
conjecture that
in the first case the marginal fixed point corresponding to $x_1=x_M$ is
selected, in the sense that as long as the initial length distribution
decays sufficiently fast with $n$, the system evolves towards the $x_1=x_M$
fixed-point. On the other hand, in the second case (the non-linear marginal
stability, or case {\uppercase \expandafter {\romannumeral2}}) the selected
fixed-point is
the one corresponding to $x_1=x_S$. In both cases the selected distribution
corresponds to the physically accessible fixed-point with the fastest decay
rate of $x_n$ with $n$.

This model has been studied numerically, and it was shown that indeed the
long time behavior of the system is governed by either the $x_1=x_M$ or the
$x_1=x_S$ fixed points depending the parameters $k, \bar k$ and $a$. One of
the parameters, say $\bar k$, may be taken as $1$. It determines the time
scale in the problem and does not affect the steady state distribution.
The resulting $(k,a)$ phase diagram is given in Fig.~2.

\bigskip
\centerline{\epsfxsize4truein\epsfbox{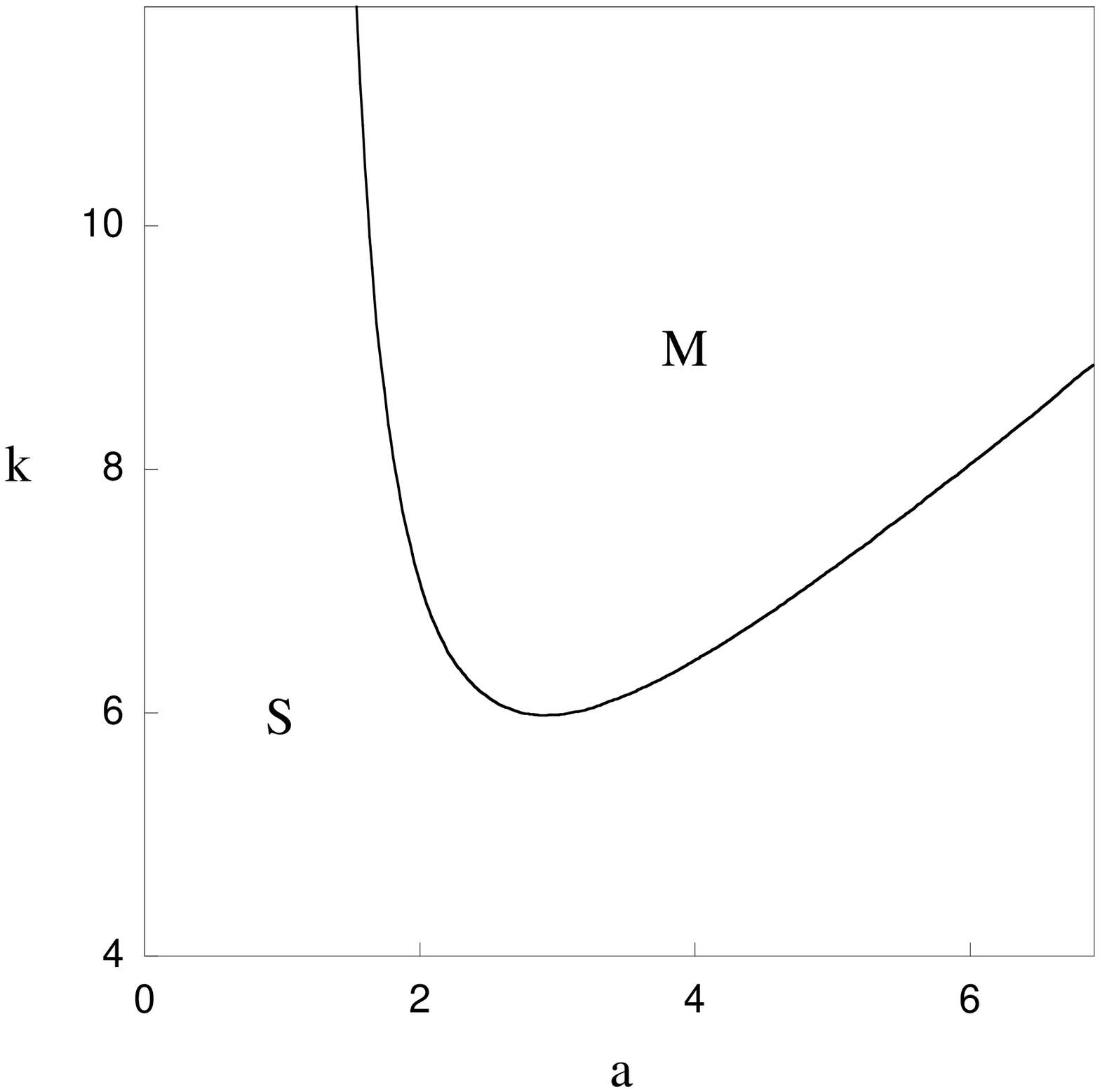}}
\nobreak\smallskip\nobreak
\offset{70pt}{Figure 2:}
The $(k,a)$ phase diagram of the model $(21)$ for $\bar k =1$. Two regions
are found. One $(M)$ in which the marginal fixed point $x_1=x_M$ is
selected, and the other, $(S)$, in which $x_1=x_S$ is selected.
\bigskip\bigskip
\noindent
{\bf Acknowledgements}

I thank E. Domany, O. Krichevsky, A. C. Maggs, C. A. Pillet, D. Segel, and
J. Stavans with whom I collaborated on the studies described in this brief
review.
\endpage
\noindent
{\bf References}
\medskip
\pointbegin
D.G. Aronson, H.F. Weinberger {\it Adv. Math.} {\bf 30}, 33 (1978).
\point
Pierre Collet and Jean-Pierre Eckmann {\it Instabilities and Fronts in
Extended Systems}, Princeton University Press, New Jersy (1990).
\point
G. Dee and J.S. Langer {\it Phys. Rev. Lett.} {\bf 50}, 383 (1983). J.S.
Langer and H. Muller-Krumbhaar {\it Phys. Rev.} {\bf A 27}, 499 (1983).
\point
E. Ben-Jacob, H. Brand, G. Dee, L. Kramer and J.S. Langer {\it Physica}
{\bf 14D}, 384 (1985).
\point
Wim van Saarloos {\it Phys. Rev. Lett.} {\bf 58}, 2571 (1987). \point
G.T. Dee and Wim van Saarloos,
{\it Phys. Rev. Lett.} {\bf 60}, 2641 (1988). \point
Wim van Saarloos {\it Phys. Rev. } {\bf A37}, 211 (1988); {\it Phys. Rev. }
{\bf A 39}, 6367 (1989). \point
Wim van Saarloos and P.C. Hohenberg, {\it Physica} {\bf D56}, 303 (1992).
\point
J. Stavans, E. Domany and D. Mukamel, {\it Europhys. Lett.} {\bf 15},
479 (1991).
\point
D. Segel, D. Mukamel, O. Krichevsky and J. Stavans, {Phys. Rev.} {\bf E
47}, 812, (1993).
\point
A.C. Maggs, D. Mukamel and C.A. Pillet, {\it Phys. Rev.} {\bf E 50}, 774
(1994).
\point
For a recent review see J. Stavans, {\it Rep. Prog. Phys.} {\bf 56}, 733
(1993).
\point
R.W. Armstrong, {\it Adv. Mater. Res.} {\bf 4}, 101 (1970). \point
C.J. Simpson, C.J. Beingessner and W.C. Winegard, {\it Trans. Metall. Soc.
AIME} {\bf 239}, 587 (1967). \point
S.K. Kurtz and F.M.A. Carpay, {J. Appl. Phys.} {\bf 51}, 5725 (1981).
\point
J.A. Glazier, S.P. Gross and J. Stavans, {\it Phys. Rev.} {\bf A 36}, 306
(1987); J. Stavans and J.A. Glazier, {\it Phys. Rev. Lett.} {\bf 62}, 1318
(1989).
\point
J. Stavans, {\it Phys. Rev.} {\bf A 42}, 5049 (1990). \point
K.L. Babcock and R.M. Westervelt, {\it Phys. Rev.} {\bf A 40}, 2022 (1989).
\point
K.L. Babcock and R.M. Westervelt, {\it Phys. Rev. Lett.} {\bf 64}, 2168 (1990).
\point
K.L. Babcock, R. Seshardri and R.M. Westervelt, {\it Phys. Rev.} {\bf A
41}, 1952 (1990).
\point
D. Weaire, F. Bolton, P. Molho and J.A. Glazier, {\it J. Phys. Condens.
Matter} {\bf 3}, 2101 (1991).
\point
J. von Neumann, in {\it Metal Interfaces}, edited by C. Herring (American
Society of Metals, Cleveland, 1952), p. 108. \point
D. Weaire and N. Rivier, {\it Contemp. Phys.} {\bf 25}, 59 (1984); N.
Rivier, {\it Philos. Mag.} {\bf B 52}, 795 (1985). \point
M.P. Marder, {\it Phys. Rev.} {\bf A 36}, 438 (1987). \point
J.R. Iglesias and R.M.C. de Almeida, {\it Phys. Rev.} {\bf A 43}, 2763 (1991).
\point
V.E. Fradkov, D.G. Udler and R.E. Kris, {\it Philos. Mag. Lett.} {\bf 58},
277 (1988).
\point
C. Beenakker, {\it Physica} {\bf A 147}, 256 (1987). \point
C. Godr\`eche, I. Kostov and I. Yekutieli, {\it Phys. Rev. Lett.} {\bf 69},
2674 (1992).
\point
B. Levitan, E. Slepyan, O. Krichevsky, J. Stavans and E. Domany,
{\it Phys. Rev. Lett.}{\bf 73}, 756 (1994).
\point
E. Br\'ezin, C. Itzykson, G. Parisi and J.-B. Zuber, {\it Comm. Math.
Phys.} {\bf 59}, 35 (1978);
D.V. Boulatov, V.A. Kazakov, I.A. Kostov and A.A. Migdal, {\it Nul. Phys.}
{\bf B275}, 641 (1986).
\endlist
\bye
\endpage
\noindent
{\bf Figure Captions}
\medskip
\offset{70pt}{Figure 1:}
$(a)$ T2 processes, describing the disappearance of 3, 4 and 5 sided cells,
respectively. $(b)$ side switching T1 process.
\medskip
\offset{70pt}{Figure 2:}
The $(k,a)$ phase diagram of the model $(21)$ for $\bar k =1$. Two regions
are found. One $(M)$ in which the marginal fixed point $x_1=x_M$ is
selected, and the other, $(S)$, in which $x_1=x_S$ is selected.

\end